\documentstyle[12pt]{article}
\newlength{\extraspace}
\newlength{\extraspaces}
\newcommand{\be}{\begin{equation}
\addtolength{\abovedisplayskip}{\extraspaces}
\addtolength{\belowdisplayskip}{\extraspaces}
\addtolength{\abovedisplayshortskip}{\extraspace}
\addtolength{\belowdisplayshortskip}{\extraspace}}
\newcommand{\ee}{\end{equation}}

\newcommand{\ba}{\begin{eqnarray}
\addtolength{\abovedisplayskip}{\extraspaces}
\addtolength{\belowdisplayskip}{\extraspaces}
\addtolength{\abovedisplayshortskip}{\extraspace}
\addtolength{\belowdisplayshortskip}{\extraspace}}
\newcommand{\ea}{\end{eqnarray}}

\newcommand{\newsection}[1]{
\vspace{15mm}
\pagebreak[3] \addtocounter{section}{1}
\setcounter{equation}{0}
\setcounter{subsection}{0}
\setcounter{footnote}{0}
\begin{flushleft}
{\large\bf \thesection. #1}
\end{flushleft}
\nopagebreak
\medskip
\nopagebreak}

\begin{document}

\addtolength{\baselineskip}{.8mm}

{\thispagestyle{empty}
\noindent \hspace{1cm} \hfill HD--THEP--00--55 \hspace{1cm}\\
\mbox{}                \hfill IFUP--TH/2000--41 \hspace{1cm}\\
\mbox{}                \hfill December 2000 \hspace{1cm}\\

\begin{center}\vspace*{0.4cm}
{\large\bf Evolution equations for the effective} \\
{\large\bf four-quark interactions in QCD} \\
\vspace*{1.0cm}
{\large Enrico Meggiolaro} \\
\vspace*{0.3cm}{\normalsize {Dipartimento di Fisica, \\ Universit\`a di
Pisa, \\ Via Buonarroti 2, \\ I-56127 Pisa, Italy.}} \\
\vspace*{0.8cm}
{\large Christof Wetterich} \\
\vspace*{0.3cm}{\normalsize {Institut f\"ur Theoretische Physik, \\
Universit\"at Heidelberg, \\ Philosophenweg 16, \\ D-69120 Heidelberg,
Germany.}} \\
\vspace*{1cm}{\large \bf Abstract}
\end{center}
\noindent
A nonperturbative renormalization group equation describes how the
momentum dependent four-quark vertex depends on an infrared cutoff.
We find a quasilocal one-particle irreducible piece generated by
(anomaly-free) multi-gluon exchange. It becomes important at a cutoff
scale where scalar and pseudoscalar meson-bound states are expected to
play a role. This interaction remains subleading as compared to the
effective one-gluon exchange contribution. The local instanton induced
four-quark interaction becomes dominant at a scale around 800 MeV.
In absence of a gluon mass the strong dependence of the one-gluon
exchange on the transferred momentum indicates that the pointlike
interactions of the Nambu-Jona-Lasinio model cannot give a very accurate
description of QCD. A pointlike effective four-quark interaction becomes
more realistic in case of spontaneous color symmetry breaking.
\vfill\eject

\newsection{Introduction}

\noindent
The Nambu-Jona-Lasinio (NJL) model \cite{A} has often been used as a
simple model for low-momentum strong interactions. After inclusion of
the chiral anomaly it gives a relatively successful description of the
chiral symmetry aspects of QCD and the associated properties of mesons
\cite{B}. A renormalization-group treatment based on exact flow
equations has provided a consistent picture of the chiral aspects of the
high temperature phase transition for two light quark flavors
\cite{C}. On the other hand it is well known that crucial properties of
long-distance QCD such as confinement cannot be accounted for by this
model. The NJL-model shares the chiral symmetries of QCD whereas color
symmetry appears only as a global symmetry. It is based on a pointlike
four-quark interaction
\be
\Gamma^{\rm (NJL)}_{\Lambda_{\rm NJL}} = \displaystyle\int {\rm d}^4x
~\left\{ i\overline{\psi}^i_a \gamma^\mu \partial_\mu \psi^a_i
+ 2 \lambda_\sigma^{\rm(NJL)} \left( \overline{\psi}_{Lb}^i \psi^a_{Ri} \right)
\left( \overline{\psi}_{Ra}^j \psi^b_{Lj} \right) \right\} ~,
\label{1.1}
\ee
where $\Lambda_{\rm NJL}$ indicates that only fluctuations with momenta
$q^2 < \Lambda_{\rm NJL}^2$ should be included in the functional
integral. (Here $i,j = 1,2,3$ and $a,b = 1,2,3$ are the color and flavor
indices, respectively.) The quartic interaction may be supplemented by a local
anomaly term which is sixth order in the quark fields for three light
flavors. Also extensions with four-quark interactions in the vector
channel have been investigated \cite{B}.

From the perspective of QCD, integrating out the gluon fields as well
as all fermion fluctuations with (covariant) momenta larger than
$\Lambda_{\rm NJL}$ should lead to an effective action containing pieces
like (\ref{1.1}), together with possibly quite complicated
additional multi-fermion interactions. The question arises under which
circumstances and for which problems the NJL model (\ref{1.1}) can be
considered as a good approximation to the multi-quark interactions
obtained from QCD. As a first step towards an answer we compute in this
paper the four-quark vertex with the index structure appearing in
Eq. (\ref{1.1}). An investigation of its momentum dependence serves as a
test if the pointlike interaction of the NJL model is reasonable.

Our method to explore the physics at the nonperturbative scale
$\Lambda_{\rm NJL}\approx (500-800)$ MeV is based on truncations of
the exact nonperturbative renormalization group equation for the
effective average action \cite{D,4a,E,6a,Ellwanger94}. An
earlier study \cite{Ellwanger-Wetterich94} has followed the running of
the momentum-dependent effective four-quark interaction with initial
conditions ``mimicking'' QCD. In contrast, we start in this paper at
short distances from standard perturbative QCD. We observe that a
one-particle-irreducible (1PI) four-quark interaction is generated by box-type
diagrams and becomes indeed approximately pointlike for scales of the order
$\Lambda_{\rm NJL}$. Nevertheless, the contribution of the one-gluon exchange
to the effective four-quark vertex retains the characteristic momentum
dependence of a particle exchange in the $t$-channel. This contribution
dominates quantitatively over the NJL-interaction. Another pointlike
interaction arises from instanton effects. It becomes dominant at a
cutoff scale around 800 MeV. We show the different contributions to the
four-quark vertex in Fig. 2, where we present the case of massless
gluons. For massless gluons a pointlike NJL-interaction of the
type (\ref{1.1}) seems not to be a realistic approximation to the effective
four-quark vertex in QCD. The hypothesis that gluons acquire a mass
$M_V \approx 850$ MeV from spontaneous color symmetry breaking \cite{F}
is discussed later (Sect. 5). For massive gluons the approximation
of pointlike four-quark interactions can be defended. Nevertheless, any
quantitatively reliable approximation should retain the one-gluon exchange
in addition to the instanton interaction and the NJL-interaction (\ref{1.1}).
This is the setting used for gluon-meson duality \cite{F}.

The paper is organized as follows. In Sect. 2 we derive a truncated
nonperturbative flow equation for the scalar-exchange-like effective
momentum-dependent four-quark interaction in QCD. This effective one-particle
irreducible interaction arises from box-type diagrams. The ``box interaction''
is relevant for the physics of scalar and pseudoscalar mesons and therefore
for spontaneous chiral symmetry breaking. In the limit where it becomes
independent of the momentum exchanged in the $t$-channel it can be
described by meson exchange. In Sect. 3 we study, by numerical methods,
the above-mentioned flow equation. We also present the perturbative expansion
of the flow equation and its solution and comment about some checks on the
validity of our approximations. In Sect. 4 we compare the box interaction
and the local instanton-induced four-quark interaction.
In Sect. 5 we discuss the hypothesis of spontaneous color symmetry
breaking, and a summary and conclusions are drawn in Sect. 6.

\newsection{Nonperturbative flow equations}

\noindent
i) EFFECTIVE AVERAGE ACTION

The effective average action $\Gamma_k$ \cite{D} generalizes the
generating functional for the 1PI correlation functions $\Gamma$
(effective action) for a situation where only fluctuations with
(covariant) momenta $q^2 > k^2$ are included in the functional integral.
It obtains by introducing an infrared cutoff $\sim k$ into the
original functional integral such that for $k\to 0$ one recovers the
effective action $\Gamma=\Gamma_{k\to 0}$. For a given $\Gamma_k$ at
nonzero $k$ the computation of $\Gamma$ from $\Gamma_k$ involves only
the fluctuations with $q^2 < k^2$. This is the setting one needs for the
effective NJL model. Our aim is therefore the computation of $\Gamma_k$
for $k=\Lambda_{\rm NJL}$. Since we do not know the precise value of
$\Lambda_{\rm NJL}$ we will consider arbitrary $k$ in the appropriate
nonperturbative range. More precisely, the NJL model is formulated in
terms of quark fields. Since we start with the QCD action involving
both quarks and gluons, the gluons should be integrated out. This may be
done by retaining the infrared cutoff only for the fermionic fields. One
then has to solve the gluon field equations obtained from $\Gamma_k$ as
a functional of the quark fields and reinsert the solution into
$\Gamma_k$ \cite{G,Wetterich96}. Typically, we will not
integrate out the gluon fluctuations with momenta smaller than $k$ in
this paper. Our results will nevertheless be a good guide for the
qualitative characteristics of the effective four-quark interactions.

The effective average action $\Gamma_k [A,c,\overline{c},
\psi,\overline{\psi}]$ for gluons $A$, ghosts $c$ and quarks $\psi$
obtains by a Legendre transform from a functional integral involving the
QCD classical action and an infrared regulator:
\be
S_k [A,c,\overline{c},\psi,\overline{\psi}] = S_{gauge} [A,c,\overline{c}] +
\Delta S_{k,gauge} [A,c,\overline{c}] + S_{F} [A,\psi,\overline{\psi}] +
\Delta S_{kF} [\psi,\overline{\psi}] ~.
\label{eq2}
\ee
Here the fermionic action $S_F$ is given by
\be
S_{F} [A,\psi,\overline{\psi}] = \displaystyle\int {\rm d}^4x~
\overline{\psi}^i_a i\gamma^\mu (D_\mu[A])_i^{\ j} \psi^a_j ~,
\label{eq3}
\ee
with $(D_\mu[A])_i^{\ j} = \delta_i^j \partial_\mu - i g A^z_\mu
(T^z)_i^{\ j}$ the covariant derivative in the fundamental representation.
Here $a$ denotes the flavor and we use two sorts of color indices
$z = 1, \ldots, N_c^2 - 1$ and $i,j = 1, \ldots, N_c$ for a gauge group
$SU(N_c)$. We consider, for simplicity, $N_f$ massless quarks: $a =
1, \ldots, N_f.$ Throughout this paper we work in four-dimensional
Euclidean space, so that the matrices $\gamma^\mu$ in the previous
equation are the (hermitean) Euclidean $\gamma$-matrices. The gauge
part $S_{gauge}$ consists of three pieces
\be
S_{gauge} [A,c,\overline{c}] = S_{YM} [A] + S_{gf} [A]
+ S_{gh} [A,c,\overline{c}] ~,
\label{eq4}
\ee
where $S_{YM}$ is the Yang-Mills action, given by:
\be
S_{YM} [A] =
{1 \over 4} \displaystyle\int {\rm d}^4x~ F^z_{\mu\nu} F^z_{\mu\nu} ~,
\label{eq5}
\ee
with
\be F^z_{\mu\nu} = \partial_\mu A^z_\nu - \partial_\nu A^z_\mu
+ g f^{zyw} A^y_\mu A^w_\nu ~.
\label{eq6}
\ee
The gauge-fixing part $S_{gf}$ reads
\be S_{gf} [A] = {1 \over 2\alpha}
\displaystyle\int {\rm d}^4x~ \partial_\mu A^z_\mu \partial_\nu A^z_\nu ~,
\label{eq7}
\ee
where $\alpha$ is the gauge parameter, and the ghost action $S_{gh}$
is given by
\be
S_{gh} [A,c,\overline{c}] = \displaystyle\int {\rm d}^4x~
\partial_\mu \overline{c}^z (\delta^{zw} \partial_\mu + gf^{zyw} A^y_\mu)
 c^w ~.
\label{eq8}
\ee
The infrared cutoff modifies the propagators. For the fermions, the
explicit expression $\Delta S_{kF} [\psi,\overline{\psi}]$ reads in
momentum space \cite{Ellwanger-Wetterich94,Wetterich96}
\be
\Delta S_{kF} [\psi,\overline{\psi}] = \displaystyle\int {{\rm d}^4q \over
(2\pi)^4}~ \overline{\psi}^i_a(q) R_{kF}(q) \psi^a_i(q) ~,
\label{eq9}
\ee
where we choose here
\ba
R_{kF}(q) &=& \gamma^\mu q_\mu
r_{kF}(q^2) ~, \nonumber \\
r_{kF}(q^2) &=& {{\rm e}^{-q^2/k^2} \over 1 - {\rm e}^{-q^2/k^2}} ~.
\label{eq10}
\ea
In particular, the (regularized) free fermion propagator turns out
to be:
\be
\left[ G^{(F)}_{2,k} (q) \right]^{ab}_{ij} = \delta^{ab}
\delta_{ij} {\gamma^\mu q_\mu \over q^2 [1 + r_{kF}(q^2)]} =
\delta^{ab} \delta_{ij} \gamma^\mu q_\mu {1 - {\rm e}^{-q^2/k^2}
\over q^2} ~.
\label{eq11}
\ee
The gauge infrared cutoff term $\Delta S_{k,gauge}
[A,c,\overline{c}]$ is the sum of a gluon infrared cutoff term and a
ghost infrared cutoff term: $\Delta S_{k,gauge} [A,c,\overline{c}] =
\Delta S_{kA} [A] + \Delta S_{k,gh} [c,\overline{c}]$. The explicit
expression of the gluon infrared cutoff term $\Delta S_{kA} [A]$ is
given by \cite{E,Ellwanger94}
\be
\Delta S_{kA} [A] = \displaystyle\frac{1}{2}\int
{{\rm d}^4p \over (2\pi)^4}~ A^z_\mu(-p) R^{kA}_{\mu\nu}(p) A^z_\nu(p) ~,
\label{eq12}
\ee
where
\be R^{kA}_{\mu\nu} (p)= \left[ \delta_{\mu\nu} + \left( {1 \over \alpha} - 1
\right) {p_\mu p_\nu \over p^2} \right] {R}_{kA}(p^2) ~,
\label{eq13}
\ee
with
\be {R}_{kA}(p^2)
 = p^2 {{\rm e}^{-p^2/k^2} \over 1 - {\rm e}^{-p^2/k^2}} ~.
\label{eq14}
\ee
Finally, the ghost infrared cutoff term
$\Delta S_{k,gh} [c,\overline{c}]$ reads
\be
\Delta S_{k,gh} [c,\overline{c}] = \displaystyle\int {{\rm d}^4p \over
(2\pi)^4}~ \overline{c}^z(-p) R_{k,gh}(p^2) c^z(p) ~,
\label{eq15}
\ee
where
\be R_{k,gh}(p^2) = p^2 {{\rm e}^{-p^2/k^2} \over 1 - {\rm e}^{-p^2/k^2}} ~.
\label{eq16}
\ee
With this choice, the regularized ghost propagator becomes
\be
\left[ G^{(gh)}_{2,k} (q) \right]^{zy} = {\delta^{zy} \over p^2 +
R_{k,gh}(p^2)} = \delta^{zy} {1 - {\rm e}^{-p^2/k^2} \over p^2} ~.
\label{eq17}
\ee

It is well known \cite{E,6a,Ellwanger94} that the presence of the infrared
cutoff explicitly breaks gauge and hence BRS invariance: in particular, a
gluon mass term appears. A background field identity \cite{E} or modified
Slavnov-Taylor identities \cite{Ellwanger94,Ellwanger-et-al.96} can be
derived, which are valid for $k > 0$. They guarantee the BRS invariance of
the theory for $k \to 0$. Within a perturbative expansion of $\Gamma_k$,
one can derive an equation for a $k$-dependent mass term for the gauge fields
implied by the modified Slavnov-Taylor identities. For pure Yang-Mills theories
(i.e., with no fermions), regularized with the above infrared cutoff term
(\ref{eq12})--(\ref{eq14}), one finds the following result, at the
leading perturbative order \cite{Ellwanger94,G}:
\be
m^2_{kA} = {3 N_c
g_k^2 \over 128 \pi^2} k^2 (\alpha - 1) ~.
\label{eq18}
\ee
Here $\alpha$ is the gauge parameter, introduced in
Eq. (\ref{eq7}), and $g_k$ is the renormalized ($k$-dependent) coupling
constant. In the theory with regularized fermions that we are
considering, a contribution to the gluon mass also appears from the
fermion part of the gluon self-energy.
One finds, at the leading perturbative order:
\be
m^2_{kF} = {3 N_f g_k^2 \over 32
\pi^2} k^2 ~.
\label{eq19}
\ee
When including the gluon mass, the regularized gluon propagator
becomes, in the Feynman gauge $\alpha = 1$:
\be
\left[ G^{(A)}_{2,k}(p) \right]^{zy}_{\mu\nu} =
{\delta_{\mu\nu} \delta^{zy} \over p^2 + m_k^2 + R_{kA}(p^2)} ~,
\label{eq20}
\ee
where
\be
m_k^2 = m^2_{kA} + m^2_{kF} ~.
\label{eq21}
\ee
(Indeed, $m^2_{kA} = 0$ at the perturbative order $O(g_k^2)$ in the
Feynman gauge.) In other words, using the expression (\ref{eq14}) for
$R_{kA}(p^2)$:
\be
\left[ G^{(A)}_{2,k}(p) \right]^{zy}_{\mu\nu} =
\delta_{\mu\nu} \delta^{zy} { 1 - {\rm e}^{-p^2/k^2} \over p^2
+ m_k^2 ( 1 - {\rm e}^{-p^2/k^2}) } ~.
\label{eq22}
\ee

We shall concentrate on the part $\Gamma^{(F)}_k [\psi,
\overline{\psi}]$ of the effective average action which depends
only on the fermion fields:
\be
\Gamma_k [A,c,\overline{c},\psi,\overline{\psi}] =
\Gamma^{(F)}_k
[\psi,\overline{\psi}] +\Gamma^{(A)}_k
[A,c,\overline{c},\psi,\overline{\psi}] ~.
\label{eq23}
\ee
In general, it can be expanded \cite{G,Wetterich96} as
\be
\Gamma^{(F)}_k [\psi,\overline{\psi}] =
\Gamma^{(F)}_{2,k} [\psi,\overline{\psi}] +
\Gamma^{(F)}_{4,k} [\psi,\overline{\psi}] + \ldots ~,
\label{eq24}
\ee
where $\Gamma^{(F)}_{2n,k} [\psi,\overline{\psi}]$ contains terms
with $2n$ fermion fields, i.e., of the form $\sim
(\overline{\psi}\psi)^n$. For $n = 1$, we have the bilinear
term\footnote{We omit in this paper the running of the fermion wave
function renormalization.}
\be
\Gamma^{(F)}_{2,k} [\psi,\overline{\psi}]
= \displaystyle\int {{\rm d}^4q \over (2\pi)^4}~
\overline{\psi}^i_a(q) \gamma^\mu q_\mu \psi^a_i(q) ~,
\label{eq25}
\ee
Similar to Refs. \cite{Ellwanger-Wetterich94,Wetterich96}, we
truncate the effective action by omitting $\Gamma^{(F)}_{2n,k}
[\psi,\overline{\psi}]$ with $n \ge 3$
and we concentrate on the behavior of the
four-quark interaction $(n=2)$. For $\Gamma^{(F)}_{4,k}
[\psi,\overline{\psi}]$ we adopt a chirally-invariant parametrization
\cite{Wetterich96}:
\ba
\lefteqn{ \Gamma^{(F)}_{4,k} [\psi,\overline{\psi}]
= - \displaystyle\int {{\rm d}^4p_1 \over (2 \pi)^4}
\displaystyle\int {{\rm d}^4p_2 \over (2 \pi)^4}
\displaystyle\int {{\rm d}^4p_3 \over (2 \pi)^4} \displaystyle\int {{\rm
d}^4p_4 \over (2 \pi)^4}~ (2\pi)^4 \delta^{(4)} (p_1+p_2-p_3-p_4) }
\nonumber \\ & & \times \left\{ \lambda_\sigma (p_1,p_2,p_3,p_4) {\cal
M}_\sigma (p_1,p_2,p_3,p_4) + \lambda_\rho (p_1,p_2,p_3,p_4) {\cal
M}_\rho (p_1,p_2,p_3,p_4) \right. \nonumber \\ & & \left. + \lambda_p
(p_1,p_2,p_3,p_4) {\cal M}_p (p_1,p_2,p_3,p_4) + \ldots \right\} ~,
\label{eq26}
\ea
where the four-fermion couplings $\lambda_i$ have the property:
\be
\lambda_i (p_1,p_2,p_3,p_4) = \lambda_i(-p_4,-p_3,-p_2,-p_1) ~,
\label{eq27}
\ee
and the usual invariant quantities $s$ and $t$ read
\be\label{2.27AA}
s = (p_1 + p_2)^2 = (p_3 + p_4)^2 ~, ~~~t = (p_1 - p_3)^2 = (p_2 - p_4)^2 ~.
\ee
The
four-fermion operators ${\cal M}_i$ are defined as:
\ba
{\cal M}_\sigma (p_1,p_2,p_3,p_4) &=&
- {1 \over 2} \left[ \overline{\psi}^i_a(-p_1)
\psi^b_i(p_2) \right] \left[ \overline{\psi}^j_b(p_4) \psi^a_j(-p_3)
\right] \nonumber \\ & & + {1 \over 2} \left[ \overline{\psi}^i_a(-p_1)
\gamma^5 \psi^b_i(p_2) \right] \left[ \overline{\psi}^j_b(p_4) \gamma^5
\psi^a_j(-p_3) \right] ~; \nonumber \\ {\cal M}_\rho (p_1,p_2,p_3,p_4)
&=& + {1 \over 4} \left[ \overline{\psi}^i_a(-p_1) \gamma^\mu
\psi^b_i(p_2) \right] \left[ \overline{\psi}^j_b(p_4) \gamma_\mu
\psi^a_j(-p_3) \right] \nonumber \\ & & + {1 \over 4} \left[
\overline{\psi}^i_a(-p_1) \gamma^\mu \gamma^5 \psi^b_i(p_2) \right]
\left[ \overline{\psi}^j_b(p_4) \gamma_\mu \gamma^5 \psi^a_j(-p_3)
\right] ~; \nonumber \\ {\cal M}_p (p_1,p_2,p_3,p_4) &=& - {1 \over 2
N_c} \left[ \overline{\psi}^i_a(-p_1) \gamma^\mu \psi^a_i(-p_3) \right]
\left[ \overline{\psi}^j_b(p_4) \gamma_\mu \psi^b_j(p_2) \right] ~.
\label{eq28}
\ea
If $\lambda_\sigma$ does not depend on $t$, the term
 ${\cal M}_\sigma$ describes the exchange of color-singlet,
flavor-nonsinglet spin-0
scalar quark-antiquark states in the $s$ channel:
they correspond to the spin-0 pseudoscalar and scalar mesons.
Similarly, ${\cal M}_\rho$ describes
the exchange of color-singlet, flavor-nonsinglet spin-1 vector
 states in the $s$ channel: they correspond to the spin-1 vector
and axial vector
mesons. Finally, ${\cal M}_p$ describes the propagation of a color-
and flavor-singlet spin-1 vector state in the $t$ channel: these are
the quantum numbers of the pomeron.
We concentrate on the analysis (both analytical
and numerical) of the evolution equation for the scalar-like effective
four-fermion coupling $\lambda_\sigma$, which multiplies ${\cal M}_\sigma$
and is relevant for the physics of scalar and pseudoscalar mesons and
therefore for spontaneous chiral symmetry breaking.
The dots in Eq. (\ref{eq26})
stand for more complicated momentum structures, the contribution of the
chiral anomaly (to be discussed in Sect. 4) and terms breaking
the chiral flavor symmetry.

\noindent
ii) FLOW EQUATIONS

The exact flow equation for $\Gamma_k$ can be written as a formal derivative
of a renormalization group improved one-loop expression \cite{D}
\be
\partial_k \Gamma_k = {1 \over 2}
{\rm STr} \left[ \tilde\partial_k \ln (\Gamma^{(2)}_k + R_k) \right] ~,
\label{eq29}
\ee
where $\Gamma^{(2)}_k$ is the second functional
derivative with respect to the fields and $\tilde\partial_k$ acts only on
$R_k$ and not on $\Gamma^{(2)}_k$, i.e.:
\be
\tilde\partial_k \equiv {\partial R_k \over \partial k} {\partial \over
\partial R_k} ~.
\label{eq30}
\ee
Here $R_k$ is to be interpreted as a matrix
$R_k = {\rm diag}~ (R_{kA},R_{k,gh},R_{kF})$.
The trace involves a summation over indices and a momentum integration
and ``STr'' indicates an additional minus sign for the fermionic part.
Flow equations for $n$-point functions follow from appropriate
functional derivatives of Eq. (\ref{eq29}) with respect to the fields.
For their derivation it is sufficient to evaluate the corresponding
one-loop expressions in presence of the
infrared cutoff (with the vertices and propagators derived from
$\Gamma_k$) and then to take a formal $\tilde\partial_k$ derivative. This
permits the use of (one-loop) Feynman diagrams and standard
perturbative techniques.

Using these techniques, we have derived the flow equation for the
four-fermion (1PI) coupling $\lambda_\sigma$ by computing the relevant
one-loop diagrams from the effective average action $\Gamma_k
[A,c,\overline{c},\psi,\overline{\psi}]$: these diagrams are shown in Fig. 1.
Our truncation includes only two vertices: i) the usual
quark-gluon vertex and ii) the four-fermion (1PI) vertex
$\lambda_\sigma {\cal M}_\sigma$. We have
extracted from the relevant diagrams the contributions to the
$\sigma$-type four-fermion vertex ${\cal M}_\sigma$ and then
performed a formal $\tilde\partial_k$ derivative in the sense of
Eq. (\ref{eq30}). In this picture (which is different from the one used in
Ref. \cite{Ellwanger-Wetterich94}) the gluons are not integrated
out. Therefore, in addition to the contributions to the flow equation
coming from the infrared cutoff in the fermion lines inside a given Feynman
diagram, one has to consider also the ones from the infrared cutoff in
the gluon lines.

\noindent
iii) TRUNCATIONS AND THE FLOW EQUATION FOR $\lambda_\sigma$

In general, the function $\lambda_\sigma$ can depend on six independent
Lorentz-invariant products of the momenta
$\lambda_\sigma = \lambda_\sigma (s,t,p_1^2,p_2^2,p_3^2,p_4^2)$.
We have derived the evolution equation of $\lambda_\sigma$ for the
particular configuration of momenta $(p_1,p_2,p_3,p_4) = (p,-p,-p,p)$,
so that $s = 0$, $t = 4 p^2$. In order to simplify the momentum structure,
we have made a further truncation, namely that $\lambda_\sigma$ depends only
on the invariant $t$: $\lambda_\sigma \simeq \lambda_\sigma (t)$.
Furthermore, our truncation restricts the quark-gluon interactions to
minimal coupling with a coupling constant depending on the gluon
momentum. We use the classical form of the (regularized) gluon
propagator\footnote{See \cite{Ellwanger-et-al.96,Bergerhoff} for an
investigation of the $q^2$-dependence of the gluon propagator by
means of nonperturbative flow equations.}.
Our truncation omits the influence of all multiquark interactions except those
proportional ${\cal M}_\sigma$ (\ref{eq28}).
Moreover, we neglect the possible effects from a $q^2$- and $k$-dependent
fermion wave-function renormalization $Z_{\psi,k}$.

With these approximations, we find the following result:
\ba
\lefteqn{ k\partial_k \lambda_\sigma (t) =
} \nonumber \\ & & = -{16 N_c \over k^2} \displaystyle\int {{\rm d}^4q
\over (2 \pi)^4} \left\{4\pi^2 \left( 1 - {3 \over 2 N_c^2} \right)
\alpha_k((p-q)^2) \alpha_k((p+q)^2) \right. \nonumber \\ &
& \times q^2 F_k(q) G_k(p-q) \left[ {\rm e}^{-q^2/k^2} G_k(p+q) +F_k(q)
\Delta_k(p+q) \right] \nonumber \\
 & &- \pi\alpha_k((p+q)^2) q^2
 \left({1 \over 2 N_c^2}
\lambda_\sigma (4p^2) \left[ {\rm
e}^{-q^2/k^2} F_k(2p+q) G_k(p+q) \right.\right. \nonumber \\
 & & \left. + {\rm
e}^{-(2p+q)^2/k^2} F_k(q) G_k(p+q) + F_k(q) F_k(2p+q) \Delta_k(p+q)
\right] \nonumber \\
& & - \left.\left( 2 - {3 \over 2 N_c^2} \right)
\lambda_\sigma ((p-q)^2) F_k(q)
 \left[ 2{\rm e}^{-q^2/k^2} G_k(p+q) + F_k(q)
\Delta_k(p+q) \right]\right) \nonumber \\
& & \Biggl. + \lambda_\sigma
((p-q)^2) \lambda_\sigma ((p+q)^2) q^2 {\rm e}^{-q^2/k^2}
F_k(q) \Biggr\} ~.
\label{eq36}
\ea
Here we use the following shorthands:
\ba
F_k(q) &=& { 1 \over q^2 [1 + r_{kF}(q^2)]} =
{ 1 - {\rm e}^{-q^2/k^2} \over q^2 } ~; \nonumber \\
G_k(q) &=& { 1 \over q^2 + m_k^2 + {R}_{kA}(q^2) }
= { 1 - {\rm e}^{-q^2/k^2} \over q^2 + m_k^2(1 - {\rm e}^{-q^2/k^2}) } ~;
\nonumber \\
\Delta_k(q) &=& { (q^2)^2 {\rm e}^{-q^2/k^2} \over
[ q^2 + m_k^2 (1 - {\rm e}^{-q^2/k^2}) ]^2 } ~,
\label{eq37}
\ea
and employ
\ba
\widetilde{\partial}_k F_k(q) &=& {\partial
r_{kF}(q^2) \over \partial k} {\partial F_k(q) \over
\partial r_{kF}(q^2)} = -{2 \over k^3} {\rm e}^{-q^2/k^2} ~,
\nonumber \\
 \widetilde{\partial}_k G_k(q) &=& {\partial
{R}_{kA}(q^2) \over \partial k} {\partial G_k(q) \over
\partial {R}_{kA}(q^2)} = -{2 \over k^3} \Delta_k(q) ~.
\label{eq39}
\ea
The first two terms in the r.h.s. of Eq. (\ref{eq36}) (i.e., those proportional
to $\sim \alpha_k^2$ or $\sim \alpha_k \lambda_\sigma$) reflect the exchange of
gauge bosons [see Fig. 1a) $\div$ 1c)].
The last term (i.e., the one proportional to $\sim \lambda_\sigma^2$)
corresponds to the result of Ref. \cite{Ellwanger-Wetterich94} for the
truncation $\lambda_\sigma = \lambda_\sigma(t)$ [see Fig. 1d)].

\newpage\noindent
iv) RG-IMPROVEMENT OF THE GAUGE COUPLING

The flow equation (\ref{eq36}) is ``RG-improved'' in the sense that we use
everywhere a ``running'' coupling constant $\alpha_k(q^2) =
g_k^2(q^2)/4\pi$. Instead of computing $\alpha_k(q^2)$ by its own flow
equation, we use here an ``ad hoc'' relation to the two-loop running
gauge coupling:
\be
\alpha(\mu)= {1 \over 4\pi\beta_0 \ln (\mu^2/\Lambda^2_{QCD}) }
\left[ 1 - {\beta_1 \over \beta_0^2} { \ln [ \ln (\mu^2/\Lambda^2_{QCD}) ]
\over \ln (\mu^2/\Lambda^2_{QCD}) } \right] ~.
\label{eq40}
\ee
Here $\beta_0$ and $\beta_1$ are the two
first coefficients of the QCD $\beta$-function
\be
\beta (g) = \mu {dg \over d\mu } = -\beta_0 g^3 -\beta_1 g^5 + \ldots ~,
\label{eq41}
\ee
which, for a gauge group $SU(N_c)$ and $N_f$ quark flavors, are given by:
\be
\beta_0 = {1 \over (4\pi)^2} \Biggl[ { 11 N_c - 2 N_f
\over 3 } \Biggr] ~,~~~ \beta_1 = {1 \over (4\pi)^4} \left[
{34 \over 3} N_c^2 - \left( { 13 N_c^2 - 3 \over 3 N_c } \right) N_f \right] ~,
\label{eq42}
\ee
such that, for $N_c = 3$ and $N_f = 3$:
\be
\alpha(\mu) = { 4\pi \over 9 \ln (\mu^2/\Lambda^2_{QCD}) }
\left[ 1 - {64 \over 81} { \ln [ \ln (\mu^2/\Lambda^2_{QCD}) ] \over
\ln (\mu^2/\Lambda^2_{QCD}) } \right] ~.
\label{eq43}
\ee
For the relation between $q^2$, $k^2$ and $\mu^2$, we use two different
{\it ans\"atze}: (A) $\mu^2 = q^2 + k^2$, and (B) $\mu^2 = k^2$.
The first one (A) is motivated by the fact that both $q^2$ and $k^2$ act as
effective infrared cutoffs such that essentially the larger one counts.
The second one (B) neglects the dependence of the vertex on the gluon momentum.
The overall mass scale needs in addition the unknown ratio between
$\Lambda_{QCD}$ in the scheme relevant for the flow equations (ERGE-scheme)
to the $\overline{\rm MS}$ scheme. We therefore report all quantities
with dimension of $[mass]^n$ in units of $({\rm GeV} \cdot \Lambda_{QCD} /
\Lambda^{(3)}_{\overline{\rm MS}})^n$, where
$\Lambda^{(3)}_{\overline{\rm MS}}$ is the fundamental mass scale of QCD
in the $\overline{\rm MS}$ renormalization scheme (two-loop), with $N_f = 3$.
The experimental value $\alpha_s(M_Z) = 0.118(3)$ corresponds to
$\Lambda^{(3)}_{\overline{\rm MS}} \simeq 360$ MeV.
For comparison we will also use a fixed value of $\alpha$
corresponding to the scale $\mu = k_c = 1.5 \cdot \Lambda_{QCD} /
\Lambda^{(3)}_{\overline{\rm MS}}$ GeV, i.e., $\alpha_{k_c} \simeq 0.34$
\quad\
[{\it ansatz} (C)].

\newpage\noindent
v) ONE-GLUON-EXCHANGE CONTRIBUTION

The term $\lambda_\sigma {\cal M}_\sigma$ describes only the contribution to
the effective $\sigma$-type vertex coming from the 1PI graphs.
A four-quark interaction with a similar structure arises also
from one-gluon exchange which is not one-particle irreducible and
therefore does not contribute\footnote{In the
formulation \cite{Ellwanger-Wetterich94,G,Wetterich96} the one-gluon
exchange part is included in $\lambda_\sigma$.} to $\lambda_\sigma$
in the present formulation. Indeed, the four-fermion vertex induced by the
one-gluon exchange is given by \cite{Wetterich96}
\ba
\displaystyle \Delta\Gamma^{(1g)}=-\int {{\rm d}^4p_1
\over (2 \pi)^4} \displaystyle\int {{\rm d}^4p_2 \over (2 \pi)^4}
\displaystyle\int {{\rm d}^4p_3 \over (2 \pi)^4} \displaystyle\int {{\rm
d}^4p_4 \over (2 \pi)^4} (2\pi)^4 \delta^{(4)} (p_1+p_2-p_3-p_4) \times
& \nonumber \\ \times 2\pi\alpha_k((p_1-p_3)^2) \cdot G_k(p_1-p_3)
\cdot {\cal M} (p_1,p_2,p_3,p_4) ~,
\label{eq48}
\ea
where
\be
{\cal M} (p_1,p_2,p_3,p_4) =
\left[ \overline{\psi}^i_a(-p_1) \gamma^\mu (T^z)_i^{\ j} \psi^a_j(-p_3)
\right] \left[ \overline{\psi}^k_b(p_4) \gamma_\mu (T_z)_k^{\ l}
\psi^b_l(p_2) \right] ~.
\label{eq49}
\ee
By an appropriate Fierz transformation, one can show that
\be
{\cal M} = {\cal M}_\sigma + {\cal M}_\rho + {\cal M}_p ~,
\label{eq50}
\ee
where ${\cal M}_\sigma$, ${\cal M}_\rho$ and ${\cal M}_p$ have been
defined in Eq. (\ref{eq28}).
The total contribution in the $\sigma$-channel is therefore
$\sim (\lambda_\sigma+\lambda_\sigma^{(1g)}){\cal M}_\sigma$ with
\be
\lambda^{(1g)}_\sigma (t) = 2\pi\alpha_k(t)
{1 - e^{-t/k^2} \over t + m^2_k (1 - e^{-t/k^2})} ~.
\label{eq51}
\ee
This contribution remains nonlocal for $k\to 0$ if $m^2_k$ tends to zero in
this limit. The size of the 1PI contribution $\lambda_\sigma$ should be
compared with the effective coupling (\ref{eq51}).

\newpage
\newsection{Scale dependence of the four-quark interaction}

\noindent
i) PERTURBATION THEORY

In perturbation theory the 1PI four-quark interactions arise
from ``box'' and ``cross'' two-gluon exchange diagrams [see Fig. 1a) and 1b)].
For the part with the structure ${\cal M}_\sigma$ we find
the perturbative value $(t=4p^2)$
\ba
\lambda_\sigma^{(p)}(t) = 16\pi^2 N_c
\left( 1 - {3 \over 2 N_c^2} \right) \displaystyle\int {{\rm d}^4q \over
(2 \pi)^4} \alpha_k((p-q)^2) \alpha_k((p+q)^2)
\nonumber \\ \times q^2 [F_k(q)]^2 G_k(p-q) G_k(p+q) ~.
\label{eq45}
\ea
In particular, for $t = 0$ and momentum-independent $\alpha_k$
[{\it ansatz} (B)], the momentum integration can be performed explicitly.
One finds, for $N_c = 3$ and in the limit of a vanishing gluon mass term
$m^2_k = 0$:
\be
\lambda_\sigma^{(p)}(t = 0) = {5 \alpha_k^2 \over k^2} \cdot K_\sigma ~.
\label{3.1AA}
\ee
Here the constant
\ba
K_\sigma &=& 8\pi^2k^2 \int \frac{{\rm d}^4q}{(2\pi)^4}
q^2 [F_k(q)]^2 [G_k^0(q)]^2 ~, \nonumber \\
G_k^0(q) &=& G_k(q) \vert_{m_k^2 = 0} = {1 \over q^2 + R_{kA}(q^2)}
= {1-e^{-q^2/k^2} \over q^2} ~,
\label{3.1BB}
\ea
depends on the precise form\footnote{For the fermion
cutoff used in \cite{4a} one has to replace
$F^2\to F/q^2$.} of the infrared cutoff. For our choice
[see Eq. (\ref{eq14})] one finds
\be
K_\sigma = 2 \cdot [5 \ln 2 - 3 \ln 3] \simeq 0.34 ~.
\label{3.1CC}
\ee
For the purpose of comparison with our results in the following figures,
we quote the value corresponding to $k = k_c = 1.5 \cdot \Lambda_{QCD} /
\Lambda^{(3)}_{\overline{\rm MS}}$ GeV ($\alpha_{k_c} \simeq 0.34)$:
\be
\lambda_\sigma^{(p)}(t = 0) \vert_{k = k_c} \simeq 8.7 \times 10^{-2}
~\left( {\rm GeV} \cdot \Lambda_{QCD} /
\Lambda^{(3)}_{\overline{\rm MS}} \right)^{-2} ~.
\label{3.1DD}
\ee

In order to get a feeling for the relative strength of the $\sigma$-channel
as compared to other types of 1PI four-quark interactions, we also present
here the full one-loop 1PI interaction for zero external momenta (i.e.,
$s = 0,~t = 0$).
For an arbitrary gauge parameter $\alpha$ one obtains the effective
(local) four-quark interaction
\ba
{\cal L}^{(F)(p)}_{4,k} &=& -\frac{g^4_k}{64\pi^2k^2}K_\sigma\nonumber\\
&&\left\{\left(\frac{9}{4}+\frac{9}{2}\alpha+\frac{3}{4}
\alpha^2\right)\left[\left(
\overline{\psi}^i_a\gamma^\mu\psi^a_j\right)\left(\overline{\psi}^j_b
\gamma_\mu\psi^b_i\right)\right.\right.
-\frac{1}{3}\left.\left(\overline{\psi}^i_a\gamma^\mu\psi_i^a\right)\left(
\overline{\psi}^j_b\gamma_\mu\psi^b_j\right)\right]\nonumber\\
&&-\frac{5}{2}\left(\overline{\psi}^i_a\gamma^\mu\gamma^5\psi^a_j\right)
\left(\overline{\psi}^j_b\gamma_\mu\gamma^5\psi^b_i\right)
\left.-\frac{11}{6}\left(\overline{\psi}^i_a\gamma^\mu\gamma^5\psi^a_i\right)
\left(\overline{\psi}^j_b\gamma_\mu\gamma^5\psi^b_j\right)\right\} ~.
\label{3.1EE}
\ea
We note that the first term has precisely the structure of the
one-gluon-exchange diagram $\sim{\cal M}$ (\ref{eq49}). We also
emphasize that there are several ways how to Fierz-transform
the expression (\ref{3.1EE}). Our truncation of keeping only
$\lambda_\sigma$ corresponds more precisely to state which
invariants are neglected.

It is straightforward to verify that the first term in the flow
equation (\ref{eq36}) for $\lambda_\sigma$ precisely corresponds
to the formal $\tilde\partial_k$ derivative of the
perturbative value (\ref{eq45}). For $t = 0$ the contribution
to $k\partial_k\lambda_\sigma$ obtains by replacing in
the r.h.s. of Eq. (\ref{3.1AA}) the constant $K_\sigma$ by
$-2 K_\sigma \simeq -0.68 ~$.

\noindent
ii) INITIAL CONDITION FOR THE FLOW EQUATION

In order to solve explicitly the flow equation (\ref{eq36}), we need to
fix the initial condition, i.e., the value of $\lambda_\sigma (t)$ at a
certain ``initial'' value $k = k_c$.
We assume that at the scale $k = k_c = 1.5 \cdot \Lambda_{QCD} /
\Lambda^{(3)}_{\overline{\rm MS}}$ GeV, where perturbation theory is
assumed to work well, the value of $\lambda_\sigma (t) \vert_{k = k_c}$
is well approximated by the leading perturbative contribution.
In other words, we use as initial condition for the flow
equation (\ref{eq36}):
\be
\lambda_\sigma(t) \vert_{k = k_c} = \lambda_\sigma^{(p)} (t) \vert_{k = k_c} ~.
\label{eq46}
\ee
We have then solved numerically the flow equation (\ref{eq36}) with
the initial condition (\ref{eq45}), (\ref{eq46}).

\noindent
iii) RESULTS

In Fig. 2 we plot the resulting $\lambda_\sigma (t)$ (for $N_c = 3$ and
$N_f = 3$) as a function of $k$, for the particular case $t = 0$. The
{\it continuous} line corresponds to
a running gauge coupling according to the {\it ansatz} (A).
The {\it dotted} line in the same figure represents
$\lambda_\sigma (t = 0)$ as a function of $k$ with the initial condition
$\lambda_\sigma (t) \vert_{k = k_c} = 0$, instead
of (\ref{eq45}), (\ref{eq46}). We observe the insensitivity with
respect to the precise choice of the initial condition.
The {\it dot-dashed} line is the value
of $\lambda_\sigma (t = 0)$ as a function of $k$ with the initial
condition (\ref{eq45}), (\ref{eq46}), but using everywhere (both in the
flow equation and in the initial condition) the momentum-independent
coupling constant $\alpha_k$ defined by Eq. (\ref{eq43}) with
$\mu^2 = k^2$ [{\it ansatz} (B)]. We observe a strong increase of
$\lambda_\sigma$ at a scale $k\approx 400-600$ MeV.

Comparison with perturbation theory is made in
Fig. 3. We plot the value of $\lambda_\sigma^{(p)} (t = 0)$,
defined by Eq. (\ref{eq45}), as a function of $k$ ({\it dotted} line).
The {\it dashed} line represents the value of $\lambda_\sigma^{(p)} (t = 0)$,
defined by Eq. (\ref{eq45}), as a function of $k$, but using the
momentum-independent coupling constant $\alpha_k$ defined by Eq.
(\ref{eq43}) with $\mu^2 = k^2$ [{\it ansatz} (B)].
In this case, the integration in Eq. (\ref{eq45}) can be performed explicitly
(neglecting also the gluon mass $m_k$ appearing in the function $G_k$) and one
obtains (for $N_c = 3$) the result (\ref{3.1AA})--(\ref{3.1CC}).
Finally, the {\it dot-dashed} line is obtained by using this
expression for $\lambda_\sigma^{(p)} (t = 0)$, but using the
fixed coupling constant $\alpha_k = \alpha_{k = k_c} \simeq 0.34$
[{\it ansatz} (C)].

In Fig. 4 we plot $\lambda_\sigma (t)$ ({\it continuous} line) as a
function of $t$ at the fixed value $k = 0.42 \cdot \Lambda_{QCD} /
\Lambda^{(3)}_{\overline{\rm MS}}$ GeV. At this scale the approximation
of the box-type four-quark interaction by a local NJL-vertex
(\ref{1.1}) is reasonable.

\noindent
iv) CHECKING THE APPROXIMATIONS

When inserting the effective action (\ref{eq23}), (\ref{eq26}) into the
flow equation (\ref{eq29}) and extracting only the contributions to the
$\sigma$-type four-fermion vertex ${\cal M}_\sigma$, one finds that in
the flow equation (\ref{eq36}) for the effective
coupling $\lambda_\sigma$ also other pieces proportional to $\sim
\lambda_i \lambda_j$ appear: these additional pieces are generated by
the four-fermion structures ${\cal M}_\rho$ and ${\cal M}_p$ and they
are put to zero in our truncation. Since only the retained contribution
$\sim \lambda_\sigma^2$ is proportional to the color factor $N_c$,
while all other terms of the form $\sim \lambda_i \lambda_j$ are
suppressed, our truncated evolution equation becomes exact in the
leading order in a $1/N_c$ expansion. Moreover, we have checked that
a multiplication of the term proportional to
$\lambda_\sigma^2$ in the flow equation (\ref{eq36}) by a factor
of ten does not strongly influence the results. The solution $\lambda_\sigma
(t)$ of the modified flow equation
is very near to the real solution of Eq. (\ref{eq36}) reported in
Figs. 2, 3 and 4. In other words, the terms of the form $\sim \lambda_i
\lambda_j$ do not seem to play a fundamental role in the flow equation
at the scales of $k$ reported here.
Neglecting the contributions from the $\rho$ and pomeron
channels does therefore not seem to be a too brutal approximation.

We have also tried to test the approximation involved in using the
two-loop expression for the running coupling constant. In order to do
this, we have considered a different version of the running coupling
constant in the flow equation (\ref{eq36}): we have taken for
$g_k^2(p^2)$ the usual two-loop expression when $g_k^2(p^2)|_{2-loop} <
10$, while $g_k^2(p^2) = 10$ when $g_k^2(p^2)|_{2-loop} > 10$. In other
words, the new $g_k^2(p^2)$ stops running when it reaches the value
$g^2_k = 10$. Also in this case, the solution $\lambda_\sigma (t)$ of
the modified flow equation does not show any dramatic difference when
compared to the solution of Eq. (\ref{eq36}) reported in Figs. 2, 3
and 4.

Finally, since the one-particle irreducible four-point function
remains small in the momentum range of interest, a neglection
of the higher-order anomaly-free effective multiquark vertices
seems well motivated.

\newsection{Comparison with the instanton contribution}

\noindent
In Eq. (\ref{eq26}) we have chosen to parametrize $\Gamma^{(F)}_{4,k}
[\psi,\overline{\psi}]$ by using $U(N_f) \otimes U(N_f)$ chirally
symmetric operators ${\cal M}_\sigma$, ${\cal M}_\rho$ and ${\cal
M}_p$. In this section we supplement
the effect of the chiral anomaly. We add to the
effective four-quark interactions an
anomaly-induced term \cite{tHooft76}, which is $SU(2)
\otimes SU(2) \otimes U(1)_V$ invariant, but not $U(1)_A$ invariant.

In particular, for a nonvanishing mass $m_s$ of the strange quark,
the instanton-induced fermion
effective action for $N_f = 3$ includes a four-fermion contribution of
the form \cite{SVZ80}
\be
\Gamma^{(4)}_{inst} = \int {\rm d}^4 x~ \lambda^{(N_c)}_{inst} {\cal
O}^{(N_c)}_{4,inst} (x) ~,
\label{eq52}
\ee
where
\ba
\lefteqn{ {\cal O}^{(N_c)}_{4,inst} = \left( \delta^j_i
\delta^l_k - {1 \over N_c} \delta^l_i \delta^j_k \right) \times }
\nonumber \\
 & & \times \left\{ [\overline{u}^i u_j] [\overline{d}^k
d_l] - [\overline{u}^i d_j] [\overline{d}^k u_l] + [\overline{u}^i
\gamma^5 u_j] [\overline{d}^k \gamma^5 d_l] - [\overline{u}^i \gamma^5
d_j] [\overline{d}^k \gamma^5 u_l] \right\} ~,
\label{eq53}
\ea
This operator comes from $\Delta L^{(N_c)}_{3,I} + \Delta
L^{(N_c)}_{3,A}$, where $I =$ instanton and $A =$ anti-instanton, and
the effective coupling constant involves an integral over instanton
sizes $\rho$:
\ba
\lambda^{(N_c)}_{inst} &=& {(1.34)^3 m_s \over 2 (N_c^2 -1) }
\displaystyle\int_0^{1/k} {{\rm d}\rho \over \rho^4}~ d^{(N_c)}_0 (\rho)
(4\pi^2\rho^3)^2 \left(\frac{\alpha(1/\rho)}{\alpha(\overline{\mu})}
\right)^{-\frac{1}{4\pi^2\beta_0}} \nonumber \\
 & = & 0.35 \cdot {m_s \over k^3} \displaystyle\int_0^{1} {\rm d}z~ z^2
\left(\frac{\alpha(k/z)}{\alpha(\overline{\mu})}\right)^{-\frac{4}{9}}
\left[ {2\pi \over \alpha(k/z)} \right]^{6} \exp \left[ -{2\pi
\over \alpha(k/z)} \right] ~.
\label{eq54}
\ea
Here $\overline{\mu} = 1$ GeV is the renormalization scale for the
fermion fields and \cite{SVZ80,Shuryak82}:
\be d^{(N_c)}_0 (\rho) =
C_{N_c} \left[ {2\pi \over \alpha(1/\rho)} \right]^{2 N_c} \exp
\left[ -{2\pi \over \alpha(1/\rho)} \right] ~,
\label{eq55}
\ee
with
\be C_{N_c} = {4.6 \exp (-1.68
N_c) \over
\pi^2 (N_c - 1)! (N_c - 2)!} ~.
\label{eq56}
\ee
For $\alpha(1/\rho)$ we take the (two-loop) running coupling
constant (\ref{eq40}) with $\mu=1/\rho$.
For the mass of the strange quark we have
used the value $m_s \simeq 150$ MeV. For $N_c = 3$ and $N_f = 3$
one has $C_3 \simeq 1.51 \times 10^{-3}$ and we use (\ref{eq43}). We
have integrated the instanton length
scale $\rho$ between $0$ and $1/k$, since $k$ is an infrared cutoff
momentum scale. The {\it long-dashed} line in Fig. 2 is the value of
$\lambda^{(3)}_{inst}$ (which is $t$-independent!) as a function of $k$.
One concludes that the instanton-induced four-quark interaction
becomes dominant at a scale $k$ around (somewhat below) 800 MeV.

\newsection{Spontaneous color symmetry breaking?}

\noindent
It has been suggested \cite{F} that confinement can be described by the
Higgs phenomenon. In this picture the expectation value of a color-octet
quark-antiquark condensate $\chi_0$ gives a mass to the gluons
\be
M^2 = g^2_{eff} \chi_0^2 ~.
\label{5.1}
\ee
In the limit of three massless quarks the mass of all eight gluons
is equal. Due to spontaneous color symmetry breaking the physical
electric charge of the gluons is integer and
they carry precisely the quantum numbers of the known vector
mesons\footnote{See \cite{F} for a discussion of the singlet vector
meson state.} $\rho,K^*,\omega$. For this reason $M$
is identified with the average mass of the vector meson octet
$M_V \approx 850$ MeV. Confinement arises in this picture in complete
analogy to magnetic flux tubes in superconductors. A nonlinear
reformulation of the Higgs phenomenon preserves manifestly
the gauge symmetry and shows the equivalence of the confinement
and Higgs pictures \cite{Dual}. Instanton-induced six-quark
interactions destabilize the ``color symmetric vacuum'' and
have been proposed as the origin of dynamical color symmetry
breaking.

The results in the preceeding sections correspond
to a vanishing octet condensate $\chi_0 = 0$. In order to get
some information on the $\chi_0$-dependence of the effective
four-quark interactions we present in this section results for a
nonvanishing value of the induced gluon mass term $M^2_k$. Since
the precise relation between $g_{eff}$ in Eq. (\ref{5.1}) and the running
gauge coupling in our approximation ({\ref{eq43})
is nontrivial \cite{F}, we present results for two
different settings.

For the first setting (A) we simply add a constant gluon mass term
$M_k^2 = M_V^2 = (850~ {\rm MeV})^2$ in the propagator such that:
\be
m_k^2 = M_V^2 + m^2_{kA} + m^2_{kF} ~.
\label{5.2}
\ee
Moreover, we use a momentum-dependent running gauge coupling $\alpha (\mu)$,
with $\mu^2 = q^2 + k^2 + M_V^2$.
For the second setting (B) we work at constant $\chi_0 = 140$ MeV with a
running perturbative gauge coupling (\ref{eq40}) $g^2_k = 4\pi\alpha$,
i.e., $M_k^2 = g^2_k \chi_0^2$. This replaces the gluon mass term in
Eq. (\ref{eq21}) by
\be
m^2_k = g^2_k \chi_0^2 + m^2_{kA} + m^2_{kF} ~.
\label{5.3}
\ee
In this case, we use a momentum-dependent running gauge coupling
$\alpha (\mu)$, with $\mu^2 = q^2 + k^2 + \overline{M}_k^2$, where
$\overline{M}_k^2 = 4 \pi \alpha(\overline{\mu} = 600 ~{\rm MeV})
\cdot \chi_0^2$.

In Figs. 5 and 6 we compare $\lambda_\sigma$ and $\lambda_\sigma^{(1g)}$ for
different running gluon mass terms $m_k^2$ given by (\ref{5.2}) (A),
(\ref{5.3}) (B), and (\ref{eq21}) (C). We observe that the gluon mass
damps the increase of both $\lambda_\sigma$ and $\lambda_\sigma^{(1g)}$.
On the other hand, a pointlike approximation to the anomaly-free
four-quark interaction becomes now better justified, especially
for (A).
Thanks to the additional infrared cutoff
$M_k^2$ the flow can formally be followed towards
$k \to 0$ without encountering a problem from a diverging gauge coupling:
the growth of the gauge coupling is reduced by the presence of the
(nonvanishing) gluon mass term.
One should not expect, however, that our results remain valid for very
small $k$. Indeed, it has been argued that the quark wave
function renormalization $Z_{\psi,k}$ multiplying the
quark kinetic term\footnote{This wave function has been set
$Z_{\psi,k}=1$ for the present work.} (\ref{eq25}) drops by a factor
of about three as $k$ reaches a few hundred MeV. This reflects the binding
of three quarks into a baryon. As a result, the running gauge coupling
$g_k$ increases by a factor of three and the renormalized
four-quark couplings $\lambda_\sigma$ and $\lambda_\sigma^{(1g)}$ become
multiplied by a factor of nine.
The running of the wave function renormalization
becomes therefore important for low $k$. We have not
included it here (except for its contribution to the
running gauge coupling), and this is the reason why we do not explore
very low values of $k$ even though the infrared problems are formally absent
in presence of a nonzero gluon mass.

We finally note that, in presence of six quark interactions reflecting
the axial anomaly, an octet condensate $\chi_0 \not= 0$ will also result
in additional effective four-quark interactions linked to the anomaly.
They are not included in the present investigation. We estimate their order
of magnitude to be comparable to (or even larger than) the
instanton contribution discussed in Sect. 4.

\newsection{Summary and conclusions}

\noindent
In this paper we have explored the effective Nambu-Jona-Lasinio-type
four-quark interactions at the nonperturbative scale
$\Lambda_{\rm NJL} \approx (500-800)$ MeV.
For this purpose we have used a method based on truncations
of the exact nonperturbative renormalization group equation for the
effective average action \cite{D,E,Ellwanger94}.
More precisely, we have derived a truncated nonperturbative flow equation for
the scalar-like effective momentum-dependent four-quark interaction in QCD.
This type of interaction is relevant for the physics of scalar and pseudoscalar
mesons and therefore for spontaneous chiral symmetry breaking.

The nonperturbative renormalization group equation describes how the
momentum dependent four-quark vertex depends on an infrared cutoff $k$.
The initial value of the vertex at short distances is computed from
perturbative QCD. We have studied the above-mentioned flow equation both
analytically, in the perturbative regime, and numerically, in the
nonperturbative regime. For the nonperturbative regime we have derived
the dependence of the scalar-like four-quark interaction on the infrared
cutoff scale $k$ and on the momentum exhanged in the $t$-channel.
The results are summarized in Figs. 2, 3 and 4.

We find that a quasilocal one-particle-irreducible (1PI) four-quark interaction
is generated and becomes indeed approximately pointlike for scales of the order
$\Lambda_{\rm NJL}$, where scalar and pseudoscalar meson-bound states are
expected to play a role. The ``quasilocality'' of this ``box-interaction'' is
put in evidence in Fig. 4, where the dependence on the momentum-scale
is shown.

As discussed in Sect. 4, another pointlike interaction arises from instanton
effects. In Fig. 2 a comparison is made between the strength of the
box interaction and that of the local instanton-induced
four-quark interaction. At scales of the order $\Lambda_{\rm NJL}$, the
instanton interaction dominates.

Nevertheless, the contribution of the one-gluon exchange to the
effective four-quark vertex retains the characteristic momentum
dependence of a particle exchange in the $t$-channel.
We show the different contributions to the four-quark vertex in
Figs. 2, 3 and 4. The effective one-gluon exchange contribution remains
substantially larger than the pointlike box-type interactions.
In consequence, the pointlike interactions (\ref{1.1}) of the
Nambu-Jona-Lasinio model cannot give a very accurate description of QCD
if the gluons are massless.
Models based on a pointlike instanton induced four-quark interaction
look more favorable. Still, the one-gluon exchange is not negligible.

The results shown in Figs. 2, 3 and 4 have been obtained with the assumption of
massless gluons (in the limit of vanishing infrared cutoff scale, $k \to 0$).
In Sect. 5 we have discussed the intriguing hypothesis that gluons acquire
a mass $M_V \approx 850$ MeV from spontaneous color symmetry breaking \cite{F}.
Our results for a nonvanishing gluon mass are shown
in Figs. 5 and 6. In particular, from Fig. 6, where the dependence on
the momentum scale is shown, it is clear that for this scenario
pointlike four-quark interactions can be defended as a much better
approximation. Nevertheless, any quantitatively reliable
approximation should retain the one-gluon exchange in addition to
the instanton interaction and the
NJL-interaction (\ref{1.1}).

We emphasize that our approximations for the flow equation in the
nonperturbative range remain relatively crude, even if in Sect. 3 we have
done some checks in order to test the validity of some of the many
approximations involved. Our results should therefore only be trusted on a
semi-quantitative level. In our opinion, they are nevertheless
a good guide for the qualitative characteristics of the effective
four-quark interactions at scales between a few hundred MeV and 1 GeV.

\vfill\eject

{\renewcommand{\Large}{\normalsize}
}

\vfill\eject

\noindent
\begin{center}
{\bf FIGURE CAPTIONS}
\end{center}
\vskip 0.5 cm
\begin{itemize}
\item [\bf Fig.~1.] One-loop diagrams for the flow equation
(\ref{eq36}) for the four-fermion (1PI) coupling $\lambda_\sigma$.
Five more diagrams similar to c) with gluon lines attached at
different quark lines are not shown.
\bigskip
\item [\bf Fig.~2.] Scale dependence of the quartic four-quark
coupling. \\
{\it Continuous} line: $\lambda_\sigma (t = 0)$ as a
function of $k$ with the initial condition as in
Eqs. (\ref{eq45}), (\ref{eq46}). {\it Dotted} line: $\lambda_\sigma (t
= 0)$ as a function of $k$ with the initial condition $\lambda_\sigma
(t) \vert_{k = k_c} = 0$ [{\it ansatz} (A)].
{\it Dot-dashed} line: the same as for
the {\it continuous} line, but using everywhere (both in the flow
equation and in the initial condition) the momentum-indepen\-dent
coupling constant $\alpha$ [{\it ansatz} (B)]. {\it Dashed} line: the
one-gluon-exchange contribution $\lambda^{(1g)}_\sigma (t = 0)$,
defined by Eq. (\ref{eq51}), as a function of $k$. {\it Long-dashed}
line: the instanton contribution $\lambda^{(3)}_{inst}$, defined by
Eq. (\ref{eq54}), as a function of $k$.
\bigskip
\item [\bf Fig.~3.] Different approximations for $\lambda_\sigma$
in perturbation theory. \\
{\it Continuous} line: the same as in Fig. 2.
{\it Dotted} line: $\lambda_\sigma^{(p)} (t = 0)$ versus $k$,
momentum-dependent $\alpha$ [{\it ansatz} (A)].
{\it Dashed} line: $\lambda_\sigma^{(p)} (t = 0)$ versus $k$,
momentum-independent $\alpha$ [{\it ansatz} (B)].
{\it Dot-dashed} line: $\lambda_\sigma^{(p)} (t = 0)$ versus $k$,
constant $\alpha$ [{\it ansatz} (C)].
\bigskip
\item [\bf Fig.~4.] Dependence of the four-quark interaction on the
transferred momentum $t$. \\
We show $\lambda_\sigma (t)$ ({\it continuous} line) and the
one-gluon-exchange contribution $\lambda^{(1g)}_\sigma (t)$ ({\it
dashed} line) as functions of $t$ at the fixed value $k = 0.42 \cdot
\Lambda_{QCD} / \Lambda^{(3)}_{\overline{\rm MS}}$ GeV.
\bigskip
\item [\bf Fig.~5.] Quartic coupling with spontaneous color symmetry
breaking. \\
{\it Continuous} and {\it dashed} lines: the same as in Fig. 2.
Lower (A) and upper (B) {\it dotted} lines: $\lambda_\sigma (t = 0)$ versus $k$
using the running gluon mass term $m_k^2$ given by Eq. (\ref{5.2}) (A)
and Eq. (\ref{5.3}) (B).
Lower (A) and upper (B) {\it dot-dashed} lines: $\lambda_\sigma^{(1g)} (t = 0)$
versus $k$ using the running gluon mass term $m_k^2$ given by Eq. (\ref{5.2})
(A) and Eq. (\ref{5.3}) (B).
\bigskip
\item [\bf Fig.~6.] Momentum dependence of quartic couplings with
spontaneous color breaking. \\
{\it Continuous} and {\it dashed} lines: the same as in Fig. 4.
Lower (A) and upper (B) {\it dotted} lines: $\lambda_\sigma (t)$ versus $t$
using the running gluon mass term $m_k^2$ given by Eq. (\ref{5.2}) (A)
and Eq. (\ref{5.3}) (B).
Lower (A) and upper (B) {\it dot-dashed} lines: $\lambda_\sigma^{(1g)} (t)$
versus $t$ using the running gluon mass term $m_k^2$ given by Eq. (\ref{5.2})
(A) and Eq. (\ref{5.3}) (B).
\end{itemize}

\vfill\eject

\pagestyle{empty}

\centerline{\bf Figure 1}
\vskip 4truecm
\begin{figure}[htb]
\vskip 4.5truecm
\includegraphics{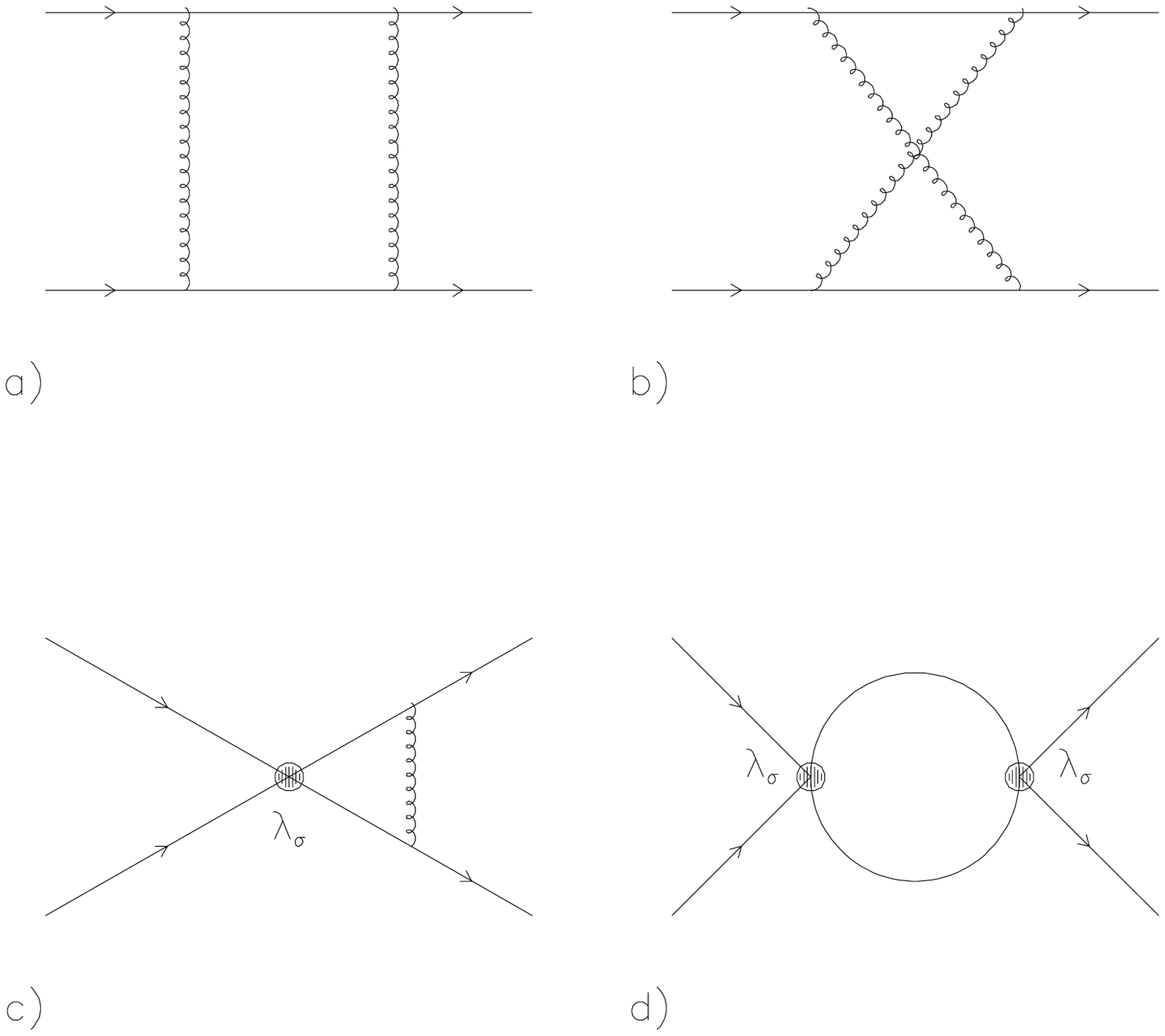}
\end{figure}

\vfill\eject

\centerline{\bf Figure 2}
\vskip 4truecm
\begin{figure}[htb]
\vskip 4.5truecm
\includegraphics{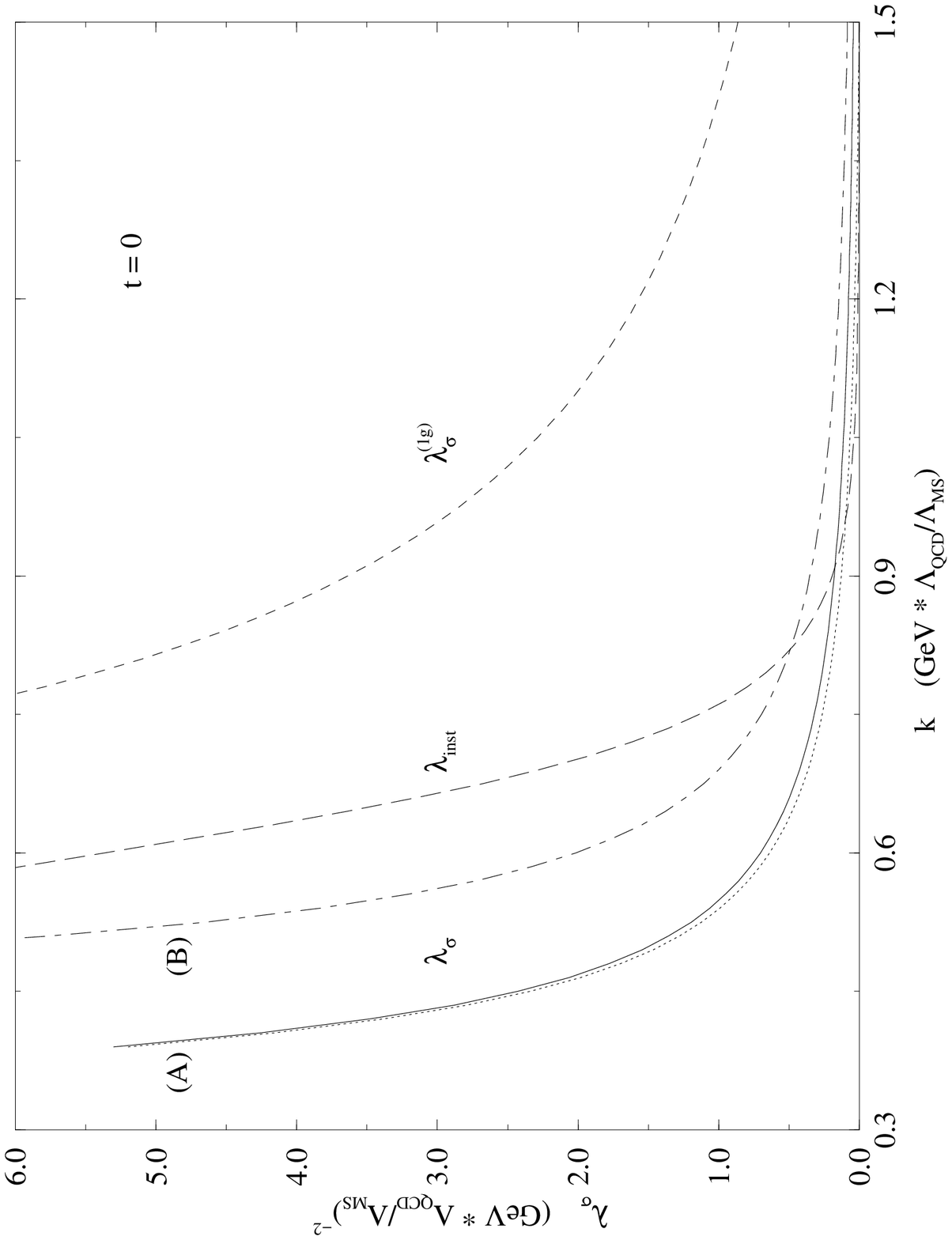}
\end{figure}

\vfill\eject

\centerline{\bf Figure 3}
\vskip 4truecm
\begin{figure}[htb]
\vskip 4.5truecm
\includegraphics{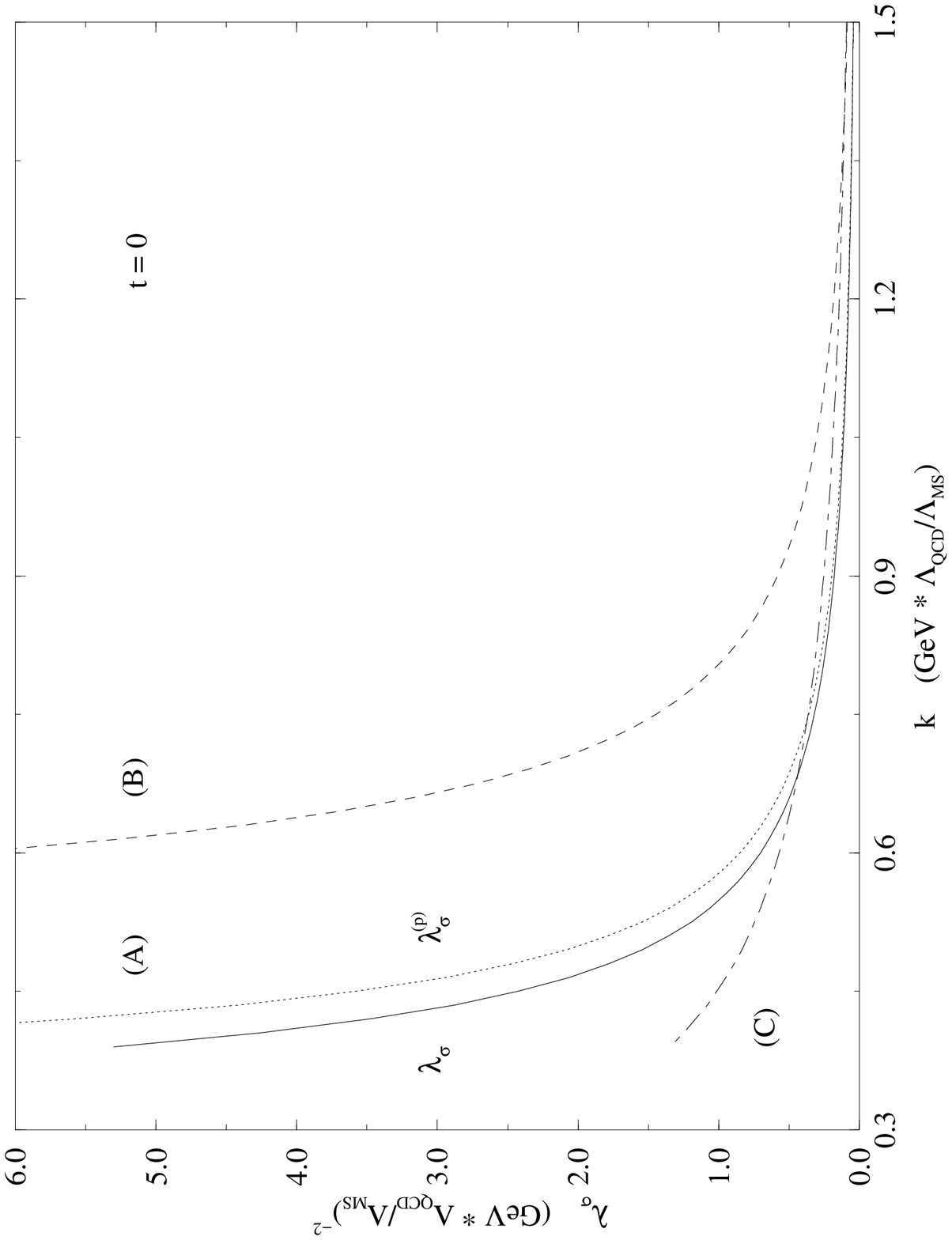}
\end{figure}

\vfill\eject

\centerline{\bf Figure 4}
\vskip 4truecm
\begin{figure}[htb]
\vskip 4.5truecm
\includegraphics{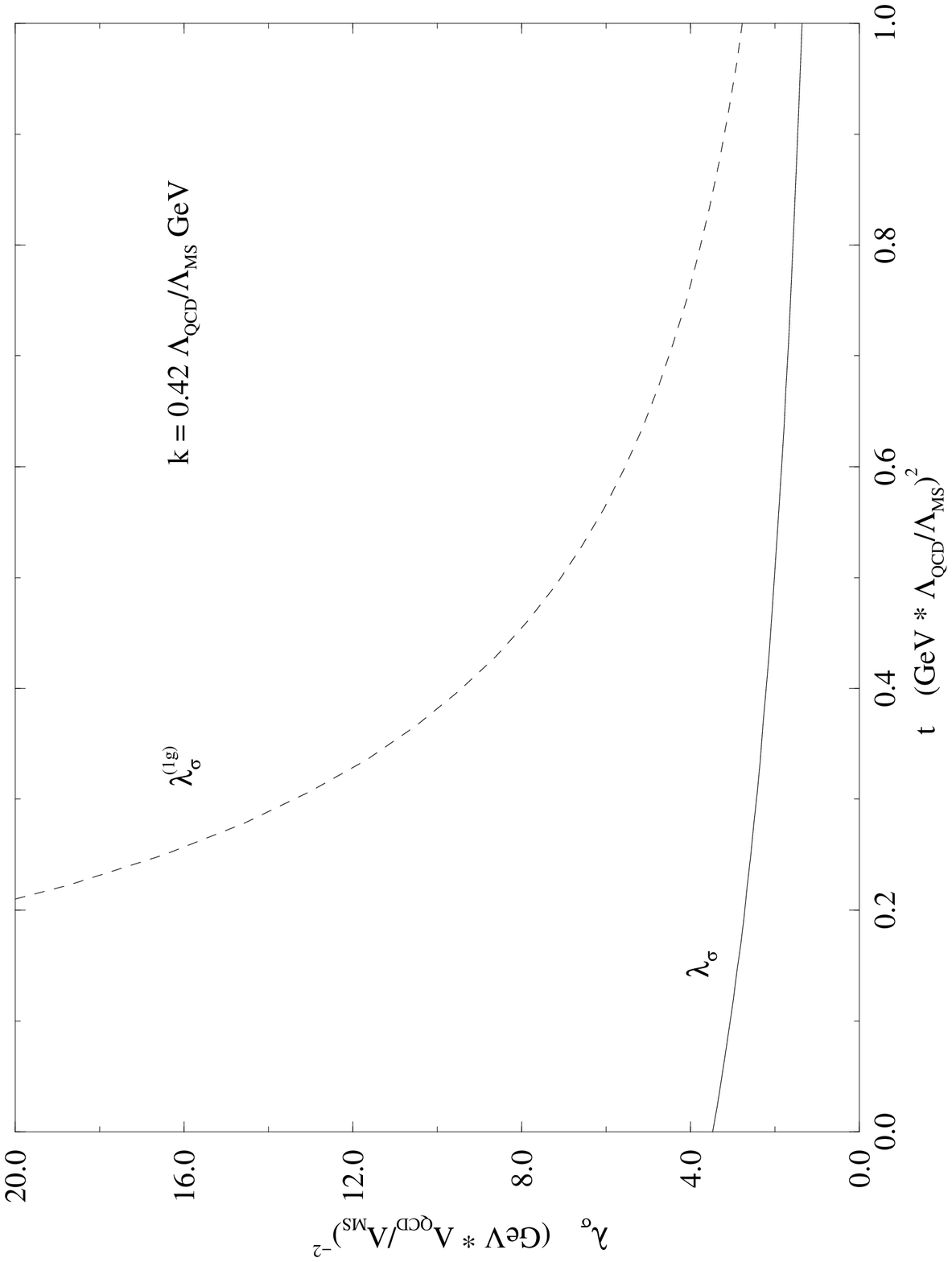}
\end{figure}

\vfill\eject

\centerline{\bf Figure 5}
\vskip 4truecm
\begin{figure}[htb]
\vskip 4.5truecm
\includegraphics{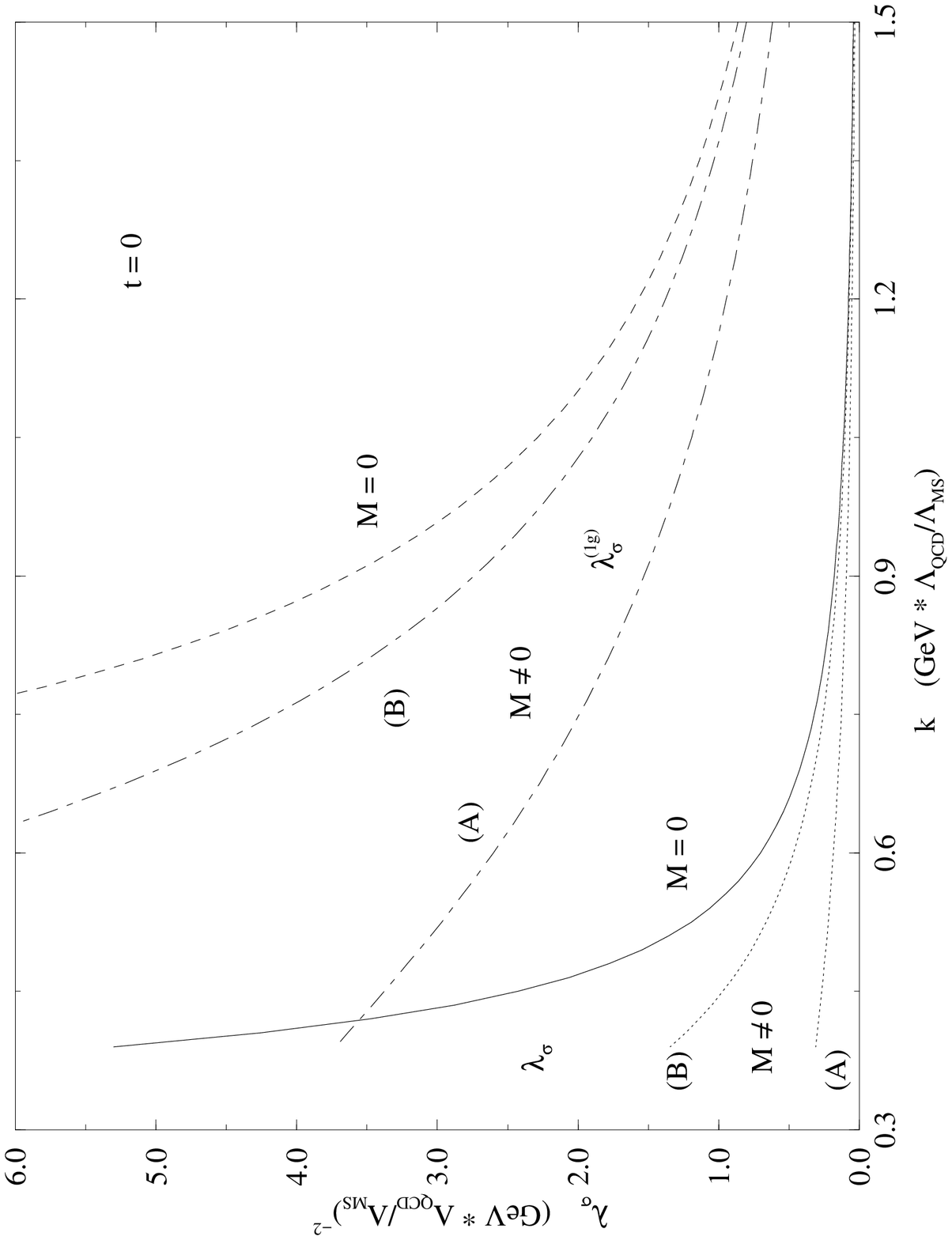}
\end{figure}

\vfill\eject

\centerline{\bf Figure 6}
\vskip 4truecm
\begin{figure}[htb]
\vskip 4.5truecm
\includegraphics{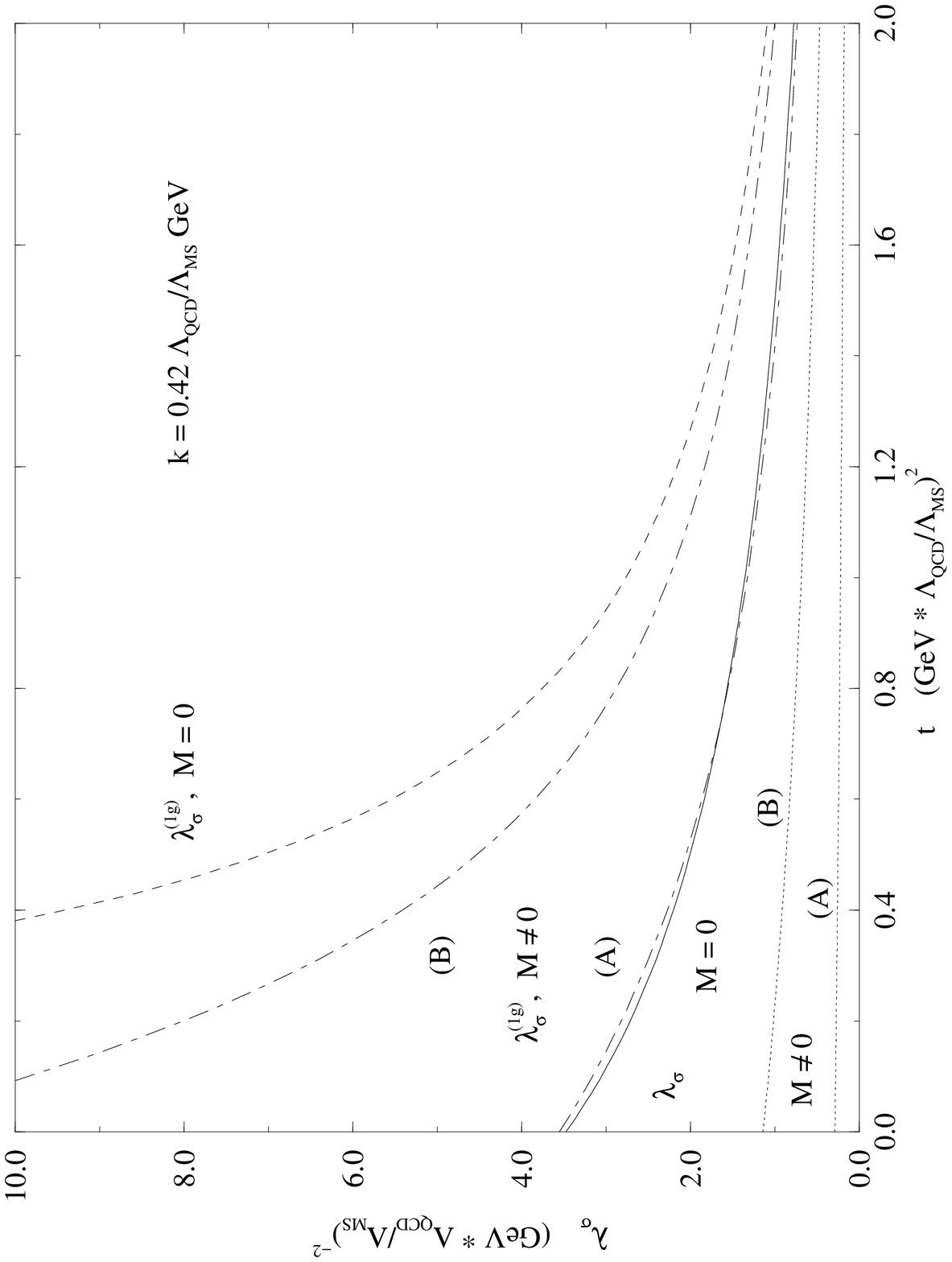}
\end{figure}

\vfill\eject

\end{document}